\documentclass[fleqn,twoside]{article}
\usepackage{espcrc2}
\usepackage{amssymb}

\usepackage{graphicx}
\usepackage{epsfig}

\newcommand{\AmS}{{\protect\the\textfont2
  A\kern-.1667em\lower.5ex\hbox{M}\kern-.125emS}}

\newcommand{\Tr}{{\mathrm {Tr}}\,}

\newcommand{\be}{\begin{equation}}
\newcommand{\ee}{\end{equation}}
\newcommand{\beqn}{\begin{eqnarray}}
\newcommand{\eeqn}{\end{eqnarray}}
\newcommand{\eq}[1]{(\ref{#1})}

\title{
\thispagestyle{empty}
\vspace{-23mm}
\rightline{\small KANAZAWA-03-16~~~~~}
\rightline{\small ITEP-LAT-2003-11~~~~~}
\vspace{10mm}
An Abelian effective action reproducing screening and confinement \newline
in quenched SU(2) QCD\thanks{Presented by K. H. at Lattice'03.}}

\author{Koichi~Hashimoto\address{Institute for Theoretical Physics, Kanazawa University,
Kanazawa 920-1192, Japan},
M. N. Chernodub${}^{\mathrm{a,}}$\address{ITEP, B.Cheremushkinskaya 25, Moscow,
117259, Russia}\thanks{M.N.Ch. is supported by JSPS Fellowship P01023.}
and Tsuneo Suzuki${}^{\mathrm{a}}$\thanks{T.S. is partially
supported by JSPS Grant-in-Aid for Scientific Research on Priority Areas
No.13135210 and (B) No.15340073. A part of our numerical simulations have been done
using NEC SX-5 at Research Center for Nuclear Physics (RCNP) of Osaka
University.}
}
\begin{document}

\begin{abstract}
In an Abelian projection $SU(2)$ gluodynamics
contains Abelian gauge fields (diagonal degrees of freedom) and
the Abelian matter fields (off-diagonal degrees). The matter fields are essential for the
breaking of the adjoint string. We obtain numerically the effective action
of Abelian fields in quenched $SU(2)$ QCD and
show that the Abelian matter fields provide an essential contribution to the total
action even in the infrared region.
\end{abstract}

\maketitle

The dual superconductor hypothesis~\cite{DualSuperconductor} was invented to
describe the confinement of color in QCD.
The model was shown to be quite successful in explanation of the confinement
of the fundamental charges such as quarks (see, $e.g.$, reviews~\cite{Reviews}).
For example, Abelian and monopole contributions
to the inter--quark potential are dominant in the long-range region of
quenched QCD~\cite{suzuki90,ref:bali}.
The monopole condensation in the confinement phase was also
observed~\cite{MonopoleCondensation}.

Although the 't~Hooft scenario describes the confinement of the
quarks correctly, this scenario predicts also the existence of the string tension for
the adjoint charges (gluons) in the infrared region. On the other hand the gluon
charges must be screened at large distances due to the presence of the gluons in
the QCD vacuum. This screening--confinement problem was discussed
within dual superconductor~\cite{SuzukiChernodub02} and center vortex
formalisms~\cite{greensite}.

The standard model of the dual superconductor in the quenched QCD
ignores the existence of the off-diagonal gluons. However, these gluons
have a charge two with respect to the Abelian subgroup and they may explain
the flattening of the inter--gluon potential which is usually studied with the help
of the adjoint Wilson loop. On the other hand the introduction of the new degrees of freedom
-- the off-diagonal gluons -- should not violate already achieved success of the
explanation of the quark confinement in this model. Indeed, quarks have
the charge one and doubly charged gluons can not screen them.

To reproduce the screening of charge two we must keep all doubly
charged Abelian Wilson loops in the effective action of the Abelian link fields.
The theory in terms of the Abelian link
fields or the Abelian monopole currents alone becomes highly non--local if we integrate
out all off-diagonal gluon fields after an Abelian projection.

Needless to say, such an
Abelian effective action is useless. The same problem is more serious in the real full QCD,
since the fundamental charge is also screened in this case.

We calculate numerically the effective action of quenched QCD
within the Abelian projection formalism. Contrary to previous calculations of this kind we include
also the doubly charged off-diagonal gluon fields into the effective action and we show that their
contribution is essential in the infrared region and thus can not be neglected.

The Wilson action of the quenched $SU(2)$ QCD is
$S = \beta \slash 2 \sum_{s,\mu,\nu} \, \Tr U_{\mu\nu}(s)$,
where $U_{\mu\nu}$ is the standard $SU(2)$ plaquette.
We parameterize the $SU(2)$ link as
$U_{\mu}(s) = c_{\mu}(s)u_{\mu}(s)$, where
$u_\mu(s) = {\mathrm{diag}}(e^{i\theta_{\mu}(s)},e^{-i\theta_{\mu}(s)})$ and
\beqn
c_\mu(s) = \left(
\begin{array}{@{}cc@{}}
\cos\phi_{\mu}(s)&
i\sin\phi_{\mu}(s)
e^{-i\varphi_{\mu}(s)}\\
i\sin\phi_{\mu}(s)
e^{i\varphi_{\mu}(s)} &
\cos\phi_{\mu}(s)
\end{array} \right).
\nonumber
\eeqn
The independent variables, $\theta$, $\varphi$, $\phi$, are restricted:
$-\pi\le\theta_{\mu}(s), \varphi_{\mu}(s)<\pi$, $0\le\phi_{\mu}(s)<\pi \slash 2$.

Under the Abelian gauge transformation,
$\Omega^{\mathrm{Abel}}(\omega) = {\mathrm{diag} (e^{i \alpha(s)},e^{- i \alpha(s)})}$,
the field $\theta$ behaves as the $U(1)$
gauge field, $\theta_{\mu}(s)\to
\theta_{\mu}(s)+\alpha(s)-\alpha(s+\hat{\mu})$,
while the field $\varphi$ corresponds to a phase of the
off--diagonal gluon field, $\varphi_{\mu}(s)\to\varphi_{\mu}(s)+2\alpha(s)$.
The variable $\phi_\mu(s)$ is not affected by the $U(1)$ gauge transformation.

If an Abelian gauge is fixed we can integrate
the $\phi_\mu(s)$ variable out without
harming the $U(1)$ content of the model.
In order to get possible forms of interactions between the Abelian gauge and Abelian matter
fields let us replace the averages of $\cos\phi_{\mu}(s)$ and $\sin\phi_{\mu}(s)$ by their
mean values in the Maximal Abelian gauge~\cite{suzukiNPBPS,Chernodub:pw}:
\beqn
\begin{array}{rcl}
\cos\phi_{\mu}(s) & \to & \langle \cos\phi_{\mu}(s) \rangle \equiv c \simeq 1\,, \\
\sin\phi_{\mu}(s) & \to & \langle \sin\phi_{\mu}(s) \rangle \equiv s \ll c\,,
\end{array}
\label{eq:mean:field}
\eeqn
Then we get an  action of the SU(2) model as
\begin{eqnarray}
\begin{array}{rcl}
\Theta_{\mu\nu}(s) \!\!\!\!& = &\!\!\!\!\theta_{\mu}(s)+\theta_{\nu}(s+\hat{\mu})
-\theta_{\mu}(s+\hat{\nu})-\theta_{\nu}(s),\\
H_{\mu\nu}(s)\!\!\!\! & = &\!\!\!\!2\theta_{\mu}(s)+\varphi_{\nu}(s)
-\varphi_{\nu}(s+\hat{\mu})\,,\\
C_{\mu\nu}(s)\!\!\!\! & = &\!\!\!\! \varphi_{\mu}(s)-\varphi_{\nu}(s)\,,
\end{array}
\nonumber
\end{eqnarray}
where $\Theta$ is the $U(1)$ plaquette for the gauge field,
$H$ describes the interaction of the matter and gauge fields,
and $C$ corresponds to the self--interaction of the matter field.
Similarly to Ref.~\cite{Chernodub:pw} we get:
\beqn
&& \frac{1}{2}\Tr U_{\mu\nu}(s) = c^4\cos(\Theta_{\mu\nu}(s))
\nonumber\\
&& +c^2s^2\cos(\Theta_{\mu\nu}(s)+C_{\mu\nu}(s)) \nonumber\\
& & + [-c^2s^2\cos(\Theta_{\mu\nu}(s)-H_{\mu\nu}(s)-C_{\mu\nu}(s)) \nonumber\\
& & + c^2s^2\cos(\Theta_{\mu\nu}(s)-H_{\mu\nu}(s)) + (\mu \leftrightarrow \nu)]\nonumber\\
&& +c^2s^2\cos(\Theta_{\mu\nu}(s)-H_{\mu\nu}(s) +H_{\nu\mu}(s)-C_{\mu\nu}(s))
\nonumber\\
& & +s^4\cos(\Theta_{\mu\nu}(s)-H_{\mu\nu}(s) +H_{\nu\mu}(s)-2C_{\mu\nu}(s)). \nonumber
\eeqn

Hence we chose, for our numerical study, a trial action
\beqn
S_{\mathrm{eff}}(\theta,\varphi) = \sum\nolimits_{i=1}^3\alpha_i S_i(\theta) + \beta_1 S_4(\theta,\varphi)\,,
\label{eq:Seff:trial}
\eeqn
where $\alpha_{i}$, $i=1,2,3$ and $\beta_1$ are the couplings
and
\beqn
&& \hspace{-5mm} S_k =-\sum\nolimits_{s,\mu\ne\nu} \cos k \Theta_{\mu\nu}(s) \,, \quad k=1,2\,,
\nonumber\\
&& \hspace{-5mm} S_3=+\sum\nolimits_{s,\mu\ne\nu} \sin\Theta_{\mu\nu}(s) \sin\Theta_{\mu\nu}(s+\hat{\mu})\,,
\label{eq:s3}\\
&& \hspace{-5mm} S_4=-\sum_{s,\mu\ne\nu} \Bigl[\cos(\Theta_{\mu\nu}(s)-H_{\mu\nu}(s))
+ (\mu \leftrightarrow \nu)\Bigr]\,.\nonumber
\eeqn
The action $S_1$ is the leading term in the Abelian action
in the mean--field approximation~\eq{eq:mean:field}. The terms $S_{2,3}$ may arise from the integration
over $\phi$ fields. The action $S_4$ describes interaction of the gauge, $\theta$, and
the matter, $\varphi$, fields.

\begin{figure}[h]
\centerline{\includegraphics[angle=-0,scale=0.35,clip=true]{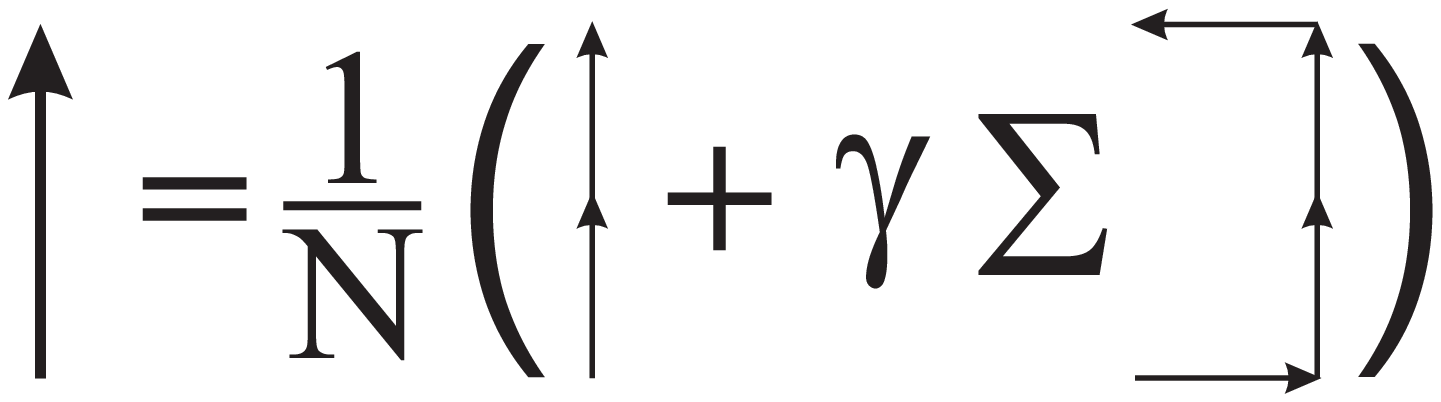}}
\vskip -7mm
\caption{The blockspin transformation.
The normalization factor, $N \equiv N(U)$, is introduced to make the fat
link belonging to the $SU(2)$ group. The weight parameter $\gamma$ is set to $1/2$.}
\label{fig:blockspin}
\end{figure}

We have used the standard Monte--Carlo procedure to generate the gauge field configurations
on the $32^4$ lattice at $\beta=2.1 \sim 2.7$.
We have generated 100 configurations of the gauge field for each value of the coupling constant
and then used the Simulated Annealing method~\cite{ref:bali} to fix the Maximal Abelian gauge.
The couplings $\alpha_i$, $i=1,2,3$ and $\beta_1$ were determined by solving the Schwinger--Dyson
equations~\cite{ref:Okawa}.
We further improve our results towards the continuum limit
using a blockspin transformation for the $SU(2)$ link variable $U_{\mu}$
visualized in Figure~\ref{fig:blockspin}.
\vskip -6mm

\begin{figure}[h]
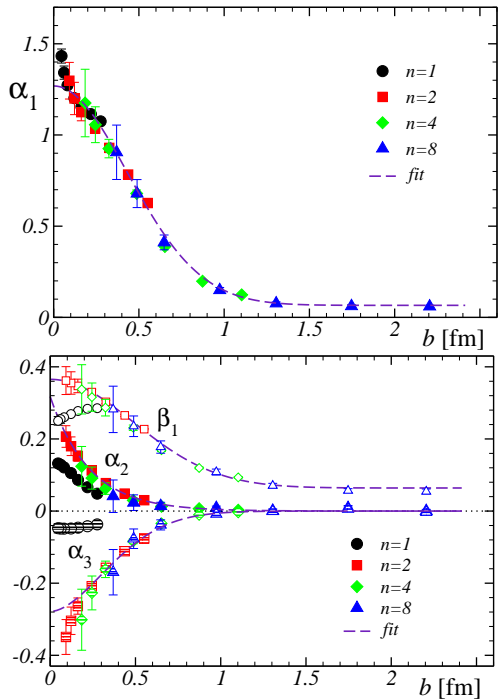

\caption{The parameters $\alpha_i$ and $\beta_1$ $vs.$ $b$.
The fits by Eq.~\eq{eq:exponential:fit} are shown by the dashed lines.}
\label{fig:alpha:beta}
\end{figure}

\vspace{-8mm}
In Figure~\ref{fig:alpha:beta} we show the couplings of the action~\eq{eq:Seff:trial} $vs.$
the scale $b=n a$ where $n$ corresponds to the blockspin factor and $a$
is the lattice spacing. The scale $b$ is shown in units of the string tension.
The coupling $\alpha_1$ shows a perfect scaling for all $n$ while the other couplings scale
for $n>1$ only. Thus at small values of $b$ the effective action
is more complicated than~\eq{eq:Seff:trial}. A similar effect is found for the
monopole action~\cite{chernodub}.

We have fitted the data for couplings by
\beqn
f(b) = C_0 + C_1 \exp\{- ({b \slash b_0)}^\nu\}\,,
\label{eq:exponential:fit}
\eeqn
where $C_{0,1}$, $\nu$ and $b_0$ are the fitting parameters. In our fits
the data with $n=1$ is excluded for all coupling constants except for $\alpha_1$.
The best fit curves are
plotted in Figures~\ref{fig:alpha:beta}, and the best fit parameters
are shown in Table~\ref{tbl:alpha:beta:fits}.

In the case of $\alpha_1$ and $\beta_1$ the parameter $\nu$ is very close to
two, therefore we fixed this parameter, $\nu=2$. Similarly, we have
set $\nu=1$ for $\alpha_2$ and $C_0=0$ for $\alpha_{2,3}$. Note that the fit
can not describe accurately the couplings $\alpha_1$ and $\alpha_3$ at small
scales, $b \leqslant 0.2$~fm, where
the Abelian action is expected to be more complicated than~\eq{eq:Seff:trial}.

\begin{center}
\includegraphics[angle=-00,scale=0.273,clip=true]{alpha1.fit.eps}
\includegraphics[angle=-0,scale=0.273,clip=true]{a23b1.fit.eps}
\end{center}
\vspace{-20mm}
\vspace{-7mm}
The functional $S_1$ makes the leading contribution to the action since
the coupling $\alpha_1$ is the largest one. The actions $S_2$ and $S_3$
play an essential role only at small distances.
The action $S_4$, which describes the interaction of the matter fields with the gauge fields is
non--vanishing at large scales similarly to $S_1$. Moreover, the
couplings $\alpha_1$ and $\beta_1$ have
larger lengths $b_0$ compared to $\alpha_2$ and $\alpha_3$.
\begin{table}
\begin{tabular}{|c|c|c|c|c|}
\hline
& $C_0$ & $C_1$ & $b_0$, [fm] & $\nu$ \\
\hline
$\alpha_1$ & 0.066(10) & 1.20(2) & 0.61(1) &2 \\
$\alpha_2$ & 0 & 0.32(2) & 0.231(7) & 1\\
$\alpha_3$ & 0 & -0.28(3) & 0.46(3) & 1.8(2) \\
$\beta_1$ & 0.064(5) & 0.30(1) & 0.69(2) & 2 \\
\hline
\end{tabular}
\caption{}
\label{tbl:alpha:beta:fits}
\vspace{-12mm}
\end{table}

Thus, at large scales, $b \sqrt{\sigma} \gg 1$, the effective Abelian action for the $SU(2)$ gauge
theory can be approximated as a sum of the QED--like action for the gauge field, $S_1(\theta)$, and
the interaction term $S_4(\theta,\varphi)$. This is the manifestation of the
Abelian dominance (non--vanishing dominant coupling $\alpha_1$) and the importance of the off-diagonal (matter)
degrees of freedom (non--vanishing coupling $\beta_1$). The matter fields are essential for the
breaking of the adjoint string.


\begin{thebibliography}{99}

\bibitem{DualSuperconductor}
G.~'t~Hooft, in {\it High Energy Physics}, ed. A. Zichichi,
EPS International Conference, Palermo (1975);
S.~Mandelstam, {\it Phys.\ Rept.}  {\bf 23} (1976) 245.

\bibitem{Reviews}
T.~Suzuki, Nucl.\ Phys.\ Proc.\ Suppl.\  {\bf 30} (1993) 176;
M.~N.~Chernodub, M.~I.~Polikarpov,
in "Confinement, duality, and nonperturbative aspects of QCD",
Ed. by P. van Baal, Plenum Press, p. 387, hep-th/9710205;
R.W. Haymaker, Phys.\ Rept.\  {\bf 315} (1999) 153.

\bibitem{suzuki90}
T.~Suzuki, I.~Yotsuyanagi,  Phys.\ Rev.\ D{\bf 42} (1990) 4257;
H.~Shiba, T.~Suzuki,
Phys.\ Lett.\ B {\bf 333} (1994) 461;
T.~Suzuki, in {\it Continuous Advances in QCD 1996}
(World Scientific, 1997),  p. 262;
J.~D.~Stack, S.~D.~Neiman and R.~J.~Wensley,
Phys.\ Rev.\ D {\bf 50} (1994) 3399.

\bibitem{ref:bali}
G.~S.~Bali {\it et al},
Phys.\ Rev.\ D {\bf 54} (1996) 2863.

\bibitem{MonopoleCondensation}
N.~Arasaki {\it et al},
Phys.\ Lett.\ B {\bf 395} (1997) 275;
K.~Yamagishi, T.~Suzuki, S.~i.~Kitahara,
JHEP {\bf 0002} (2000) 012;
M.~N.~Chernodub, M.~I.~Polikarpov, A.~I.~Veselov,
Phys.\ Lett.\ B {\bf 399}, 267 (1997);
Nucl.\ Phys.\ Proc.\ Suppl.\  {\bf 49} (1996) 307;
A.~Di Giacomo, G.~Paffuti,
Phys.\ Rev.\ D {\bf 56} (1997) 6816;
H.~Shiba, T.~Suzuki,  Phys.\ Lett.\ B{\bf 351} (1995) 519.

\bibitem{SuzukiChernodub02}
T.~Suzuki and M.~N.~Chernodub,
Phys.\ Lett.\ B {\bf 563} (2003) 183.

\bibitem{greensite}
L.~Del Debbio {\it et al},
Phys.\ Rev.\ D {\bf 58} (1998) 094501;
J.~Ambj\o rn  {\it et al}, JHEP  {\bf 0002} (2000) 033.

\bibitem{suzukiNPBPS}
T.~Suzuki and I.~Yotsuyanagi,
Nucl.\ Phys.\ Proc.\ Suppl.\  {\bf 20} (1991) 236.

\bibitem{Chernodub:pw} M.~N.~Chernodub, M.~I.~Polikarpov and A.~I.~Veselov,
Phys.\ Lett.\ B {\bf 342} (1995) 303.

\bibitem{ref:Okawa}
A.~Gonzalez-Arroyo, M.~Okawa,
Phys.\ Rev.\ D {\bf 35} (1987) 672;
Phys.\ Rev.\ B {\bf 35} (1987) 2108.

\bibitem{chernodub}
M.~N.~Chernodub,
{\it et al},
Phys.\ Rev.\ D {\bf 62}, 094506 (2000);
hep-lat/9902013.

\end{thebibliography}
\end{document}